\documentclass[authoryear,5p]{elsarticle}

\usepackage{graphicx}
\usepackage{amssymb}
\usepackage{amsmath}

\usepackage{lineno}

\usepackage{multirow}
\usepackage{url}

\def\nuc#1#2{${}^{\mathrm{#1}}$#2}

\begin{document}

\begin{frontmatter}

\title{Arrival time and magnitude of airborne fission products from the Fukushima, Japan, reactor incident as measured in Seattle, WA, USA}

\author[UW]{J.~Diaz~Leon}
\author[UWBoth]{D.\,A.~ Jaffe}
\author[UW]{J.~Kaspar}
\author[UW]{A.~Knecht\corref{cor_auth}}\ead{knechta@uw.edu}
\author[UW]{M.\,L.~Miller}
\author[UW]{R.\,G.\,H.~Robertson}
\author[UW]{A.\,G.~Schubert}

\address[UW]{Department of Physics and Center for Experimental Nuclear Physics and Astrophysics, University of Washington, Seattle, WA 98195, USA}
\address[UWBoth]{Department of Interdisciplinary Arts and Sciences, University of Washington, Bothell, WA 98011, USA}

\cortext[cor_auth]{Corresponding author}

\begin{abstract}We report results of air monitoring started due to the recent natural catastrophe on 11 March 2011 in Japan and the severe ensuing damage to the Fukushima Dai-ichi nuclear reactor complex. On 17-18 March 2011, we registered the first arrival of the airborne fission products  \nuc{131}{I}, \nuc{132}{I}, \nuc{132}{Te}, \nuc{134}{Cs}, and \nuc{137}{Cs} in Seattle, WA, USA, by identifying their characteristic gamma rays using a germanium detector. We measured the evolution of the activities over a period of 23 days at the end of which the activities had mostly fallen below our detection limit. The highest detected activity from radionuclides attached to particulate matter amounted to 4.4$\pm$1.3~mBq\,m$^{-3}$ of \nuc{131}{I} on 19-20 March.
\end{abstract}

\end{frontmatter}

\section{Introduction}
The recent earthquake and tsunami in Japan on 11 March 2011, 05:46 Coordinated Universal Time (UTC) resulted in severe damage to the Fukushima Dai-ichi nuclear reactor complex. Due to the uncertainty of the situation, limited quantitative information, and its potential impact on both local public health as well as our low-background fundamental physics program \citep{Ell09}, we began monitoring local air samples in Seattle, WA, USA, for the potential arrival of airborne radioactive fission products. In this paper we present data on key radionuclides associated with the nuclear accident and a brief discussion of the transport. We believe it is important to provide a rapid release of this data for two reasons:
\begin{itemize}
\item While the earthquake, tsunami and nuclear accident are extremely serious within Japan, there has also been a great deal of concern around nuclear radiation reaching the U.S.  Our data, generated by independent university research, should provide greater confidence in our understanding of the risks from long-distance transport of radionuclides.  
\item These data, along with other observations, will help provide better information on the characteristics of the source, the release mechanism, and the transport of the radionuclides. 
\end{itemize} 

\section{Material and Methods}
Our samples consist of air filters taken from the intake to the ventilation system of the Physics and Astronomy building at the University of Washington. This allows us to sample $\sim$10 times more air than what had been done previously here at the university after the Chernobyl incident \citep{Kel86} and proved to be one of the key points for the successful detection of the radioactive fission products. In order to search for characteristic gamma rays stemming from radionuclides we place the samples inside a lead shield of 5 to 20~cm thickness next to a \mbox{0.5-kg} p-type point contact germanium detector \citep{Bar07} for low-level counting. The detector exhibits an energy resolution of 1.4~keV FWHM at 600~keV. The level of observed background radiation inside the shield ranges from 10~counts/keV/hour at 50~keV to 2~counts/keV/hour at 800~keV. The energy and efficiency of the detector have been calibrated using 10 strong gamma lines between 200 and 1500~keV from a \nuc{152}{Eu} source to an accuracy of about 0.1~keV and 10\%, respectively.

We digitize the preamplified traces coming from the germanium detector using the Struck card SIS3302 \citep{Struck} which at the same time extracts the energy of the measured pulse. The communication with the card and the VME crate is managed by ORCA \citep{How04}. After the acquisition the data are automatically uploaded to a database for analysis. In addition to the real physics events we also inject a pulser signal into the preamplifier to check the live time and health of the system. The pulser runs at 0.1~Hz with an amplitude equivalent to an energy of $\sim$120~keV. The signal from the pulser is clearly visible in Fig.~\ref{fullSpectrum}.

The air filters used are commercial ventilation filters from AmericanAirFilter (Model PerfectPleat ULTRA)  and Purolator (Model DMK80-STD2) with dimensions 61$\times$61$\times$5~cm. Their efficiency for retaining particles down to a size of 5~$\mu$m amounts to 75\%, drops to 35\% at 1~$\mu$m and to 5\% at 0.4~$\mu$m. From our detection of the cosmogenic \nuc{7}{Be} isotope, we calculated an activity of $\sim$0.1~mBq\,m$^{-3}$. Comparing this value to the known \nuc{7}{Be} concentration of 2-8~mBq\,m$^{-3}$ \citep{Yos03} we deduce a filter efficiency of 2.4$\pm$2.0\% or, correspondingly, particle sizes of $\lesssim$0.4~$\mu$m. This roughly agrees with observations of radioactive particle sizes after the Chernobyl accident \citep{Dev86,Jos86} and measured sizes of radioactive dust in the atmosphere \citep{Ber73}. In an auxiliary measurement we exposed in addition to our standard filter a high efficiency particulate air filter (Hunter HEPAtech filter model 30930). This filter is $\sim$100\% efficient for particles of sizes $\gtrsim$0.1~$\mu$m and features one layer of activated charcoal. Taking the measured activities from that filter as the true activities in air, we calculate the efficiency for our standard filters from the ratio of measured activities. The filter efficiency amounts to 14.2$\pm$3.2\% for the \nuc{131}{I} and 4.4$\pm$1.2\% for the other fission products and \nuc{7}{Be}, in agreement with the above estimate. This points to slightly larger particle sizes of \nuc{131}{I} compared to the other isotopes ($\sim$0.6~$\mu$m instead of $\sim$0.4~$\mu$m given the filter efficiencies). We can only speculate on the initial conditions and the chemistry involved that would lead to this difference. A difference between the particle sizes for \nuc{131}{I} and other fission products had already been observed after the Chernobyl accident \citep{Jos86}. However, in those measurements the \nuc{131}{I} particle sizes tended to be slightly smaller than the other fission products. We believe that the difference between those measurements and our findings is due to the longer transport distance. The long transport tends to lead to overall smaller particle sizes. As \nuc{131}{I} can be transported part of the way in its elemental form and can thus be adsorbed at later stages, this could lead to larger \nuc{131}{I} particle sizes with respect to the other fission products which are adsorbed shortly after the release. 

At several other radiation monitoring stations cartridges filled with charcoal were used in addition to particle air filters \citep{Radnet, Tay}. The charcoal captures the \nuc{131}{I} present in its gaseous form and the combination of the two thus measures the total amount of \nuc{131}{I} in air. The ratio of gaseous \nuc{131}{I} to \nuc{131}{I} attached to particulate matter varied between 2 and 20 with an average of about 5. We would like to stress that our results below only measure the portion of \nuc{131}{I} attached to particulate matter.

The air filters were typically exposed for one day to an air flow of 114000$ \pm$8000~m$^3$/day, which was measured using a Davis 271 Turbo-Meter flowmeter. We bagged and compressed the filters into packages of approximately 1000 to 4000~cm$^{3}$ before placing them into the lead shield for counting. The solid angle for gamma rays emitted within that volume and interacting with the germanium detector was calculated given the actual dimensions and ranged between 1.6 and 4.4\%. We attribute a 20\% systematic uncertainty to the calculated value.

\section{Results and Discussion}

\begin{figure}[t]
\vspace*{2mm}
\begin{center}
\includegraphics[width=1.0\linewidth, angle=0]{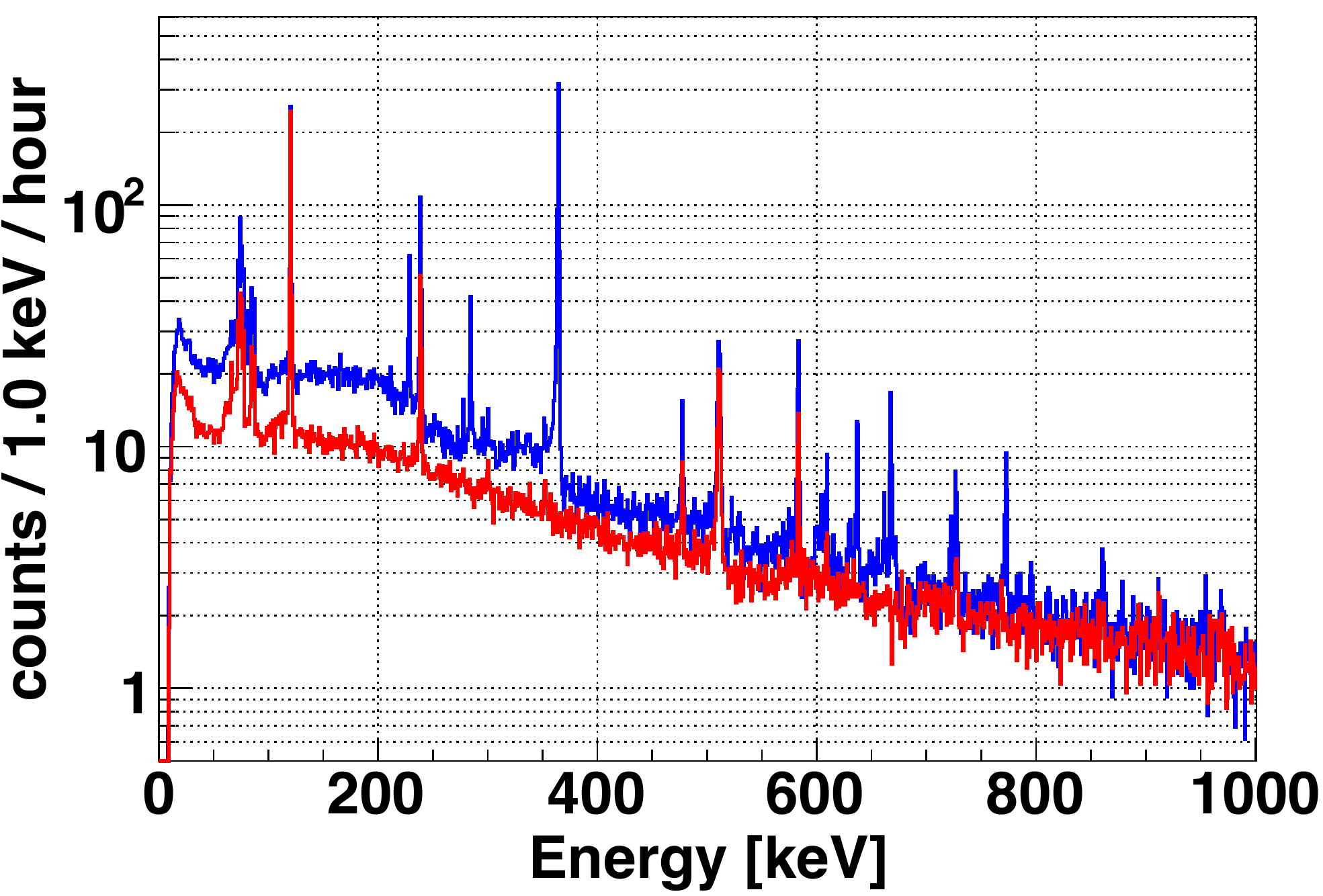}
\end{center}
\caption{\label{fullSpectrum}(Colour online) Comparison of the gamma spectra from the measurements of air filter PH1 (red, 16-17 March) and air filter PH2 (blue, 17-18 March) showing clearly the additional peaks due to the arrival of radioactive fission products at the US west coast. The dominant peak at 364 keV is from \nuc{131}{I}.}
\end{figure}

\begin{figure}[t]
\vspace*{2mm}
\begin{center}
\includegraphics[width=1.0\linewidth, angle=0]{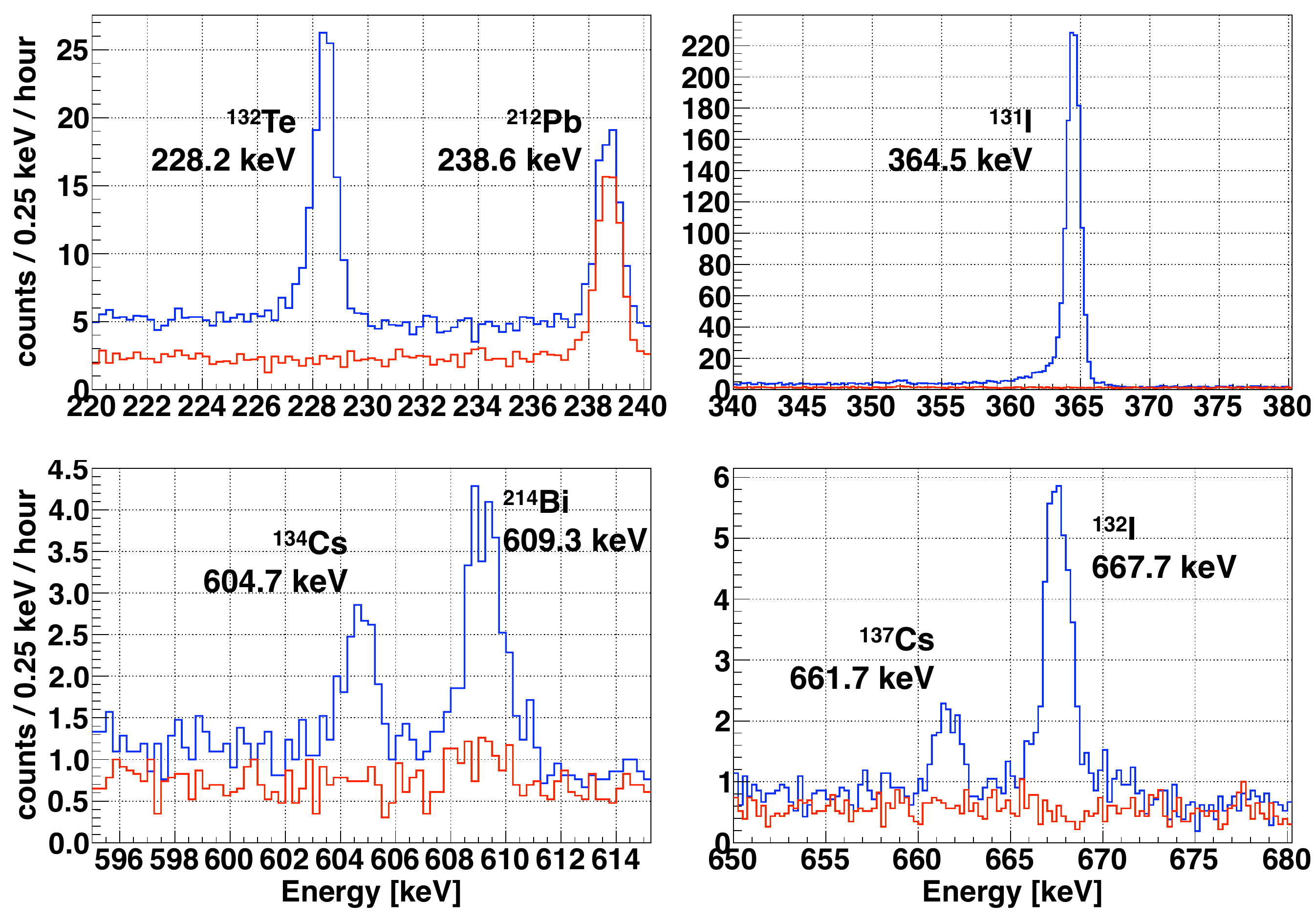}
\end{center}
\caption{\label{zoomedSpectrum}(Colour online) Plot of the 5 strongest gamma lines of \nuc{131}{I}, \nuc{132}{I}, \nuc{132}{Te}, \nuc{134}{Cs}, and \nuc{137}{Cs} for the air filter PH1 (red) and air filter PH4 (blue) measurements. The change in \nuc{214}{Bi}activity is due to fluctuating radon levels during the time of measurement.}
\end{figure}

We started the air monitoring campaign on 16 March 2011. The exact exposure and counting periods for the different air filters are listed in Table~\ref{filter_tab}. No fission products were detected in the first air filter PH1 and we were able to attribute all the visible gamma lines to known background radioactivity from cosmic-ray induced processes, various radioactive isotopes of the uranium and thorium decay chains, cosmogenic \nuc{7}{Be}, and \nuc{40}{K}. The subsequent sample PH2 immediately revealed the onset of several characteristic gamma lines from fission products. The identified isotopes are \nuc{131}{I}, \nuc{132}{I}, \nuc{132}{Te}, \nuc{134}{Cs}, and \nuc{137}{Cs}. Figure~\ref{fullSpectrum} shows the comparison between the gamma ray spectra from air filter PH1 and PH2 where the additional gamma peaks are clearly identifiable. Figure~\ref{zoomedSpectrum} demonstrates the statistical significance of the detected lines by showing the peaks of the strongest decay branches of the five identified isotopes. 

\begin{figure}[t]
\vspace*{2mm}
\begin{center}
\includegraphics[width=1.0\linewidth, angle=0]{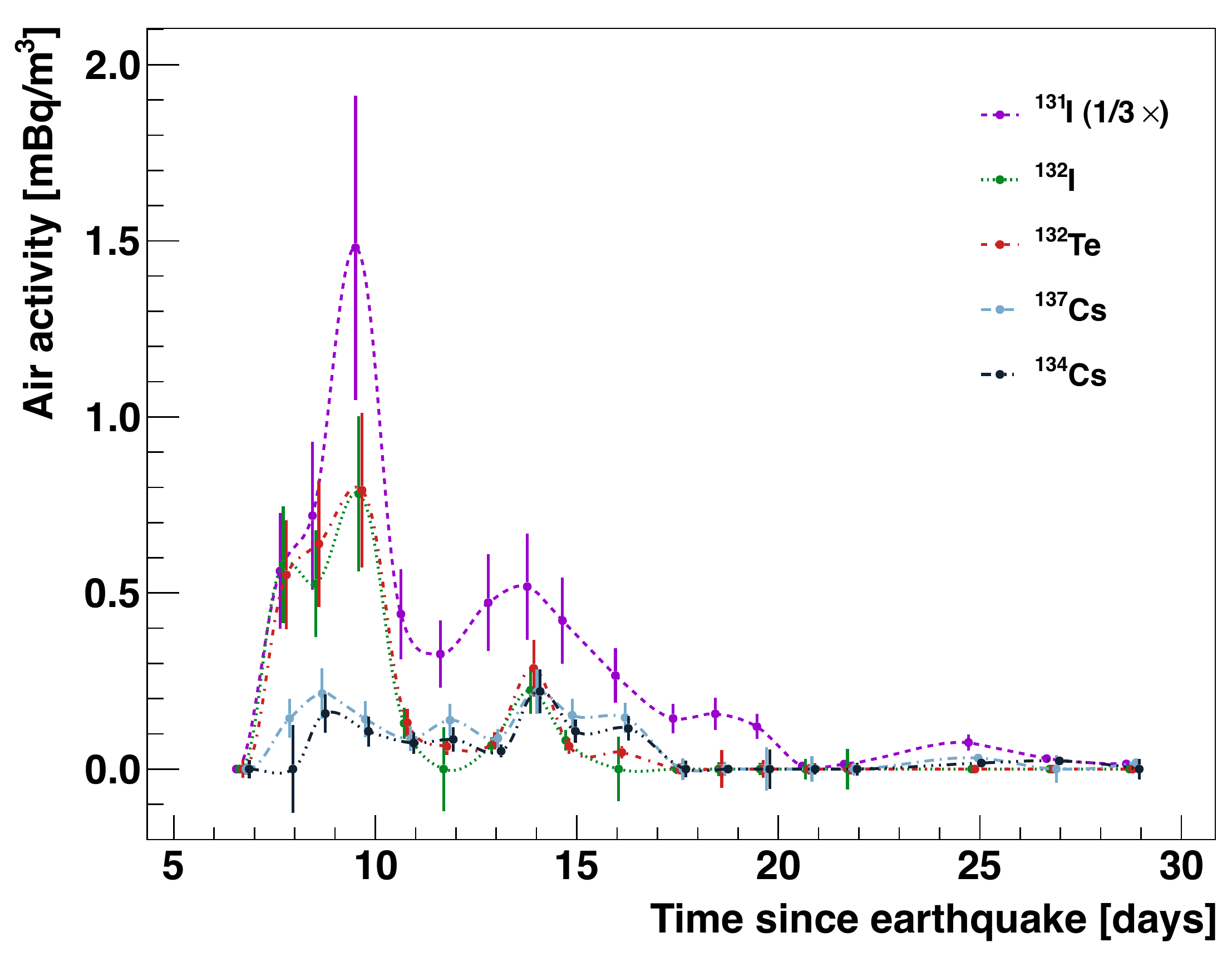}
\end{center}
\caption{\label{airActivities}(Colour online) Activities of the five fission isotopes \nuc{131}{I}, \nuc{132}{I}, \nuc{132}{Te}, \nuc{134}{Cs}, and \nuc{137}{Cs} in air for the exposed air filters covering a period of 23 days. Results for \nuc{131}{I} are scaled by a factor 1/3 in order to improve the visibility of the other radionuclides. The exact values can be found in Table~\ref{activity_tab}. The errors show the statistical and systematic uncertainties added in quadrature. Points are shown with small horizontal offsets to increase visibility.}
\end{figure}

In order to obtain the decay rates of the different isotopes we summed the appropriate spectral bins in the region of interest and subtracted the sum of the same amount of bins in the side bands. For the cases in which the statistical significance of the extracted signal counts was less than three sigma we calculated the upper limit at 95\% C.L. by employing the Feldman-Cousins formalism \citep{Fel98} for a possible signal given the number of background counts. From the detected number of counts ($N_{\mathrm{det}}$ ) and the given counting time ($T$) and isotope lifetime ($\tau$) we calculate the decay rate ($R_{\mathrm{det}}$) at the beginning of the counting period:
\begin{equation}
R_{\mathrm{det}} = \frac{N_{\mathrm{det}}}{\tau \left(1 -  e^{-T/\tau} \right)}
\end{equation}
From this and measurements and calculations of air flow ($Q$), filter, geometrical, and detection efficiency ($\epsilon_{\mathrm{filter}}$, $\epsilon_{\mathrm{geom}}$ and $\epsilon_{\mathrm{det}}$) the amount of activity present in air ($A_{\mathrm{air}}$) using the known branching ratios ($\mathrm{BR}$) is: 
\begin{equation}
A_{\mathrm{air}} = \frac{R_{\mathrm{det}}}{\mathrm{BR} \epsilon_{\mathrm{det}} \epsilon_{\mathrm{geom}}} \frac{1}{Q \epsilon_{\mathrm{filter}}}
\end{equation}
In addition, we correct the values for delays in the counting period ($t_{\mathrm{del}}$) with respect to the end of the exposure using the known lifetimes ($\tau$) thus calculating the activity present in the filter at the end of the exposure time:
\begin{equation}
A'_{\mathrm{air}} = A_{\mathrm{air}} e^{t_{\mathrm{del}}/\tau}
\end{equation}
We do not correct for decays during the exposure itself as the time structure of the arrival of the radioactive atoms is unknown. The values for the activities can be found in Fig.~\ref{airActivities} and Table~\ref{activity_tab}. We cross-checked for consistency between the activities obtained from different branches of the same isotope. As they were in agreement and the systematic error dominates we only give the value for the strongest branch. For \nuc{132}{I} with a half life of only 2.3 hours we do not detect the atoms originating in the reactor but the ones following the decay of \nuc{132}{Te}. The measured activities for these two isotopes given in Table~\ref{activity_tab} agree very nicely corroborating our correction due to the energy dependent detection efficiency. The highest observed activity of \nuc{131}{I} amounts to 4.4$\pm$1.3~mBq\,m$^{-3}$.

\begin{table*}[t]
%\scriptsize
\caption{\label{filter_tab}Exposure and counting periods for the different air filters. Times are given in UTC. Days after earthquake give the time at the end of the exposure.}
\vskip4mm
\centering
\begin{tabular}{l  c   r @{ -- } l c  r @{ -- } l }
\hline
Filter & Brand & \multicolumn{2}{c}{Exposure}& Days after earthquake & \multicolumn{2}{c}{Counting}  \\
\hline
PH1 & AmericanAir & 16/3 17:00 &  17/3 19:00 & 6.6 & 17/3 23:45 &  18/3 22:45\\
PH2 & AmericanAir & 17/3 19:00 &  18/3 21:00 & 7.6 & 18/3 23:30 & 19/3 02:30\\
\multicolumn{5}{c}{} & 19/3 06:00 & 19/3 16:00\\
PH3 & AmericanAir & 18/3 21:00 &  19/3 16:20 & 8.4 & 19/3 18:30 &  20/3 18:30\\
PH4 & Purolator & 19/3 16:20 &  20/3 18:00 & 9.5 & 20/3 19:30 &  21/3 16:30\\
PH5 & Purolator & 20/3 18:00 &  21/3 21:00 & 10.6 & 21/3 23:20 &  22/3 21:05\\
PH6 & Purolator & 21/3 21:00 & 22/3 20:30 & 11.6 &  22/3 21:16 & 23/3 02:16 \\
\multicolumn{5}{c}{} & 24/3 01:44 & 24/3 20:45\\
PH7 & Purolator & 22/3 20:30 & 24/3 01:00 & 12.8 & 24/3 22:48 & 25/3 21:48 \\
PH8 & Purolator & 24/3 01:00 & 25/3 00:15 & 13.8 & 27/3 05:08 & 27/3 20:08 \\
PH9 & Purolator & 25/3 00:15 & 25/3 21:00 & 14.6 & 26/3 17:59 & 27/3 04:59 \\
PH10 & Purolator & 25/3 21:00 & 27/3 04:45 & 16.0 & 28/3 16:11 & 29/3 01:11 \\
PH11 & Purolator & 27/3 04:45 & 28/3 15:05 & 17.4 & 29/3 01:48 & 29/3 15:48 \\
PH12 & Purolator & 28/3 15:05 & 29/3 16:15 & 18.4 & 29/3 16:59 & 30/3 19:00 \\
PH13 & Purolator & 29/3 16:15 & 30/3 17:00  & 19.5 & 30/3 20:27 & 01/4 02:27 \\
PH14 & Purolator & 30/3 17:30 & 31/3 20:00  & 20.6 & 01/4 18:14 & 03/4 01:14 \\
PH15 & Purolator & 31/3 20:00 & 01/4 21:00  & 21.6 & 04/4 15:37 & 05/4 18:14 \\
PH16 & Purolator & 01/4 21:00 & 04/4 23:00  & 24.7 & 05/4 19:05 & 07/4 17:05 \\
PH17 & Purolator & 04/4 23:00 & 06/4 21:30  & 26.7 & 07/4 20:22 & 08/4 19:22 \\
PH18 & Purolator & 06/4 21:30 & 08/4 21:00  & 28.6 & 10/4 03:35 & 11/4 22:35 \\
\hline
\end{tabular}
\end{table*}

\begin{table*}[t]
%\tiny
\scriptsize
%\footnotesize
\caption{\label{activity_tab}Detected activities and upper limits for the measured radionuclides attached to particulate matter. The numbers are given in $\mu$Bq per m$^3$. The errors give the 1$\sigma$ values for the systematic and statistical uncertainties (in that order). In case of upper limits we give the limit from statistics only at 95\% C.L. and quote the influence of the systematic error separately at 1$\sigma$. The upper limits strongly depend on the fluctuations of background radiation at the time of counting.
}
\vskip4mm
\centering
\begin{tabular}{@{ }lr@{}lr@{}lr@{}lr@{}lr@{}lr@{}lr@{}l@{ }} 
\hline
 & \multicolumn{2}{l}{\nuc{7}{Be}} & \multicolumn{2}{l}{\nuc{131}{I}} & \multicolumn{2}{l}{\nuc{132}{I}} & \multicolumn{2}{l}{\nuc{132}{Te}} & \multicolumn{2}{l}{\nuc{133}{I}} & \multicolumn{2}{l}{\nuc{134}{Cs}} & \multicolumn{2}{l}{\nuc{137}{Cs}} \\ 
Sample & \multicolumn{2}{l}{53.2~days}& \multicolumn{2}{l}{8.0~days}& \multicolumn{2}{l}{2.3~hours}& \multicolumn{2}{l}{3.2~days}& \multicolumn{2}{l}{20.8~hours}& \multicolumn{2}{l}{2.1~years}& \multicolumn{2}{l}{30.1~years}\\ 
 & \multicolumn{2}{l}{477.6~keV}& \multicolumn{2}{l}{364.5~keV}& \multicolumn{2}{l}{667.7~keV}& \multicolumn{2}{l}{228.2~keV}& \multicolumn{2}{l}{529.9~keV}& \multicolumn{2}{l}{604.7~keV}& \multicolumn{2}{l}{661.7~keV}\\ 
\hline
PH1
 & 1200~ & $\pm$~300~$\pm$~200  % 7-Be 478 
 & \textless~2.5~ & $\pm$~0.7  % 131-I 364 
 & \textless~16~ & $\pm$~5  % 132-Te 228 
 & \textless~6~ & $\pm$~2  % 132-I 668 
 & \textless~70~ & $\pm$~20  % 133-I 530 
 & \textless~17~ & $\pm$~5  % 134-Cs 605 
 & \textless~11~ & $\pm$~3  % 137-Cs 662 
\\ 
PH2
 & 2200~ & $\pm$~600~$\pm$~300  % 7-Be 478 
 & 1690~ & $\pm$~490~$\pm$~20  % 131-I 364 
 & 550~ & $\pm$~150~$\pm$~30  % 132-Te 228 
 & 580~ & $\pm$~160~$\pm$~50  % 132-I 668 
 & \textless~170~ & $\pm$~50  % 133-I 530 
 & \textless~80~ & $\pm$~20  % 134-Cs 605 
 & 140~ & $\pm$~40~$\pm$~40  % 137-Cs 662 
\\ 
PH3
 & 4100~ & $\pm$~1100~$\pm$~300  % 7-Be 478 
 & 2160~ & $\pm$~630~$\pm$~20  % 131-I 364 
 & 640~ & $\pm$~180~$\pm$~30  % 132-Te 228 
 & 530~ & $\pm$~150~$\pm$~40  % 132-I 668 
 & \textless~110~ & $\pm$~30  % 133-I 530 
 & 160~ & $\pm$~40~$\pm$~30  % 134-Cs 605 
 & 210~ & $\pm$~60~$\pm$~40  % 137-Cs 662 
\\ 
PH4
 & 1400~ & $\pm$~400~$\pm$~200  % 7-Be 478 
 & 4440~ & $\pm$~1300~$\pm$~30  % 131-I 364 
 & 790~ & $\pm$~220~$\pm$~30  % 132-Te 228 
 & 780~ & $\pm$~220~$\pm$~40  % 132-I 668 
 & \textless~60~ & $\pm$~20  % 133-I 530 
 & 110~ & $\pm$~30~$\pm$~30  % 134-Cs 605 
 & 140~ & $\pm$~40~$\pm$~30  % 137-Cs 662 
\\ 
PH5
 & 1100~ & $\pm$~300~$\pm$~200  % 7-Be 478 
 & 1320~ & $\pm$~380~$\pm$~20  % 131-I 364 
 & 130~ & $\pm$~40~$\pm$~20  % 132-Te 228 
 & 130~ & $\pm$~40~$\pm$~30  % 132-I 668 
 & \textless~40~ & $\pm$~10  % 133-I 530 
 & 70~ & $\pm$~20~$\pm$~20  % 134-Cs 605 
 & 90~ & $\pm$~20~$\pm$~30  % 137-Cs 662 
\\ 
PH6
 & 1200~ & $\pm$~300~$\pm$~200  % 7-Be 478 
 & 980~ & $\pm$~290~$\pm$~10  % 131-I 364 
 & 60~ & $\pm$~20~$\pm$~20  % 132-Te 228 
 & \textless~80~ & $\pm$~20  % 132-I 668 
 & \textless~9~ & $\pm$~3  % 133-I 530 
 & 80~ & $\pm$~20~$\pm$~30  % 134-Cs 605 
 & 140~ & $\pm$~40~$\pm$~30  % 137-Cs 662 
\\ 
PH7
 & 1560~ & $\pm$~430~$\pm$~100  % 7-Be 478 
 & 1190~ & $\pm$~350~$\pm$~10  % 131-I 364 
 & \textless~50~ & $\pm$~10  % 132-Te 228 
 & \textless~70~ & $\pm$~20  % 132-I 668 
 & \textless~180~ & $\pm$~50  % 133-I 530 
 & 39~ & $\pm$~11~$\pm$~9  % 134-Cs 605 
 & \textless~50~ & $\pm$~10  % 137-Cs 662 
\\ 
PH8
 & 1900~ & $\pm$~500~$\pm$~100  % 7-Be 478 
 & 1550~ & $\pm$~450~$\pm$~20  % 131-I 364 
 & 290~ & $\pm$~80~$\pm$~20  % 132-Te 228 
 & 220~ & $\pm$~60~$\pm$~30  % 132-I 668 
 & \textless~170~ & $\pm$~50  % 133-I 530 
 & 220~ & $\pm$~60~$\pm$~20  % 134-Cs 605 
 & 220~ & $\pm$~60~$\pm$~20  % 137-Cs 662 
\\ 
PH9
 & 2100~ & $\pm$~600~$\pm$~200  % 7-Be 478 
 & 1260~ & $\pm$~370~$\pm$~20  % 131-I 364 
 & 70~ & $\pm$~20~$\pm$~10  % 132-Te 228 
 & 80~ & $\pm$~20~$\pm$~20  % 132-I 668 
 & \textless~80~ & $\pm$~20  % 133-I 530 
 & 110~ & $\pm$~30~$\pm$~20  % 134-Cs 605 
 & 150~ & $\pm$~40~$\pm$~20  % 137-Cs 662 
\\ 
PH10
 & 1900~ & $\pm$~500~$\pm$~100  % 7-Be 478 
 & 800~ & $\pm$~230~$\pm$~10  % 131-I 364 
 & 50~ & $\pm$~10~$\pm$~10  % 132-Te 228 
 & \textless~60~ & $\pm$~20  % 132-I 668 
 & \textless~32~ & $\pm$~9  % 133-I 530 
 & 120~ & $\pm$~30~$\pm$~10  % 134-Cs 605 
 & 150~ & $\pm$~40~$\pm$~20  % 137-Cs 662 
\\ 
PH11
 & 1070~ & $\pm$~290~$\pm$~90  % 7-Be 478 
 & 431~ & $\pm$~126~$\pm$~7  % 131-I 364 
 & \textless~10~ & $\pm$~3  % 132-Te 228 
 & \textless~9~ & $\pm$~3  % 132-I 668 
 & \textless~11~ & $\pm$~3  % 133-I 530 
 & \textless~16~ & $\pm$~4  % 134-Cs 605 
 & \textless~20~ & $\pm$~5  % 137-Cs 662 
\\ 
PH12
 & 1070~ & $\pm$~300~$\pm$~100  % 7-Be 478 
 & 470~ & $\pm$~139~$\pm$~7  % 131-I 364 
 & \textless~35~ & $\pm$~10  % 132-Te 228 
 & \textless~13~ & $\pm$~4  % 132-I 668 
 & \textless~9~ & $\pm$~3  % 133-I 530 
 & \textless~0.8~ & $\pm$~0.2  % 134-Cs 605 
 & \textless~13~ & $\pm$~4  % 137-Cs 662 
\\ 
PH13
 & 1910~ & $\pm$~530~$\pm$~100  % 7-Be 478 
 & 362~ & $\pm$~107~$\pm$~6  % 131-I 364 
 & \textless~16~ & $\pm$~5  % 132-Te 228 
 & \textless~11~ & $\pm$~3  % 132-I 668 
 & \textless~6~ & $\pm$~2  % 133-I 530 
 & \textless~40~ & $\pm$~10  % 134-Cs 605 
 & \textless~40~ & $\pm$~10  % 137-Cs 662 
\\ 
PH14
 & 410~ & $\pm$~110~$\pm$~60  % 7-Be 478 
 & 26~ & $\pm$~8~$\pm$~2  % 131-I 364 
 & \textless~10~ & $\pm$~3  % 132-Te 228 
 & \textless~19~ & $\pm$~5  % 132-I 668 
 & \textless~40~ & $\pm$~10  % 133-I 530 
 & \textless~10~ & $\pm$~3  % 134-Cs 605 
 & \textless~23~ & $\pm$~6  % 137-Cs 662 
\\ 
PH15
 & 730~ & $\pm$~210~$\pm$~80  % 7-Be 478 
 & 43~ & $\pm$~13~$\pm$~3  % 131-I 364 
 & \textless~10~ & $\pm$~3  % 132-Te 228 
 & \textless~40~ & $\pm$~10  % 132-I 668 
 & \textless~270~ & $\pm$~80  % 133-I 530 
 & \textless~12~ & $\pm$~3  % 134-Cs 605 
 & \textless~12~ & $\pm$~3  % 137-Cs 662 
\\ 
PH16
 & 840~ & $\pm$~240~$\pm$~30  % 7-Be 478 
 & 227~ & $\pm$~67~$\pm$~2  % 131-I 364 
 & \textless~1.3~ & $\pm$~0.4  % 132-Te 228 
 & \textless~6~ & $\pm$~2  % 132-I 668 
 & \textless~11~ & $\pm$~3  % 133-I 530 
 & 17~ & $\pm$~5~$\pm$~3  % 134-Cs 605 
 & 31~ & $\pm$~9~$\pm$~4  % 137-Cs 662 
\\ 
PH17
 & 700~ & $\pm$~200~$\pm$~60  % 7-Be 478 
 & 89~ & $\pm$~27~$\pm$~3  % 131-I 364 
 & \textless~7~ & $\pm$~2  % 132-Te 228 
 & \textless~7~ & $\pm$~2  % 132-I 668 
 & \textless~22~ & $\pm$~6  % 133-I 530 
 & 23~ & $\pm$~7~$\pm$~7  % 134-Cs 605 
 & \textless~26~ & $\pm$~7  % 137-Cs 662 
\\ 
PH18
 & 460~ & $\pm$~130~$\pm$~40  % 7-Be 478 
 & 44~ & $\pm$~13~$\pm$~2  % 131-I 364 
 & \textless~2.4~ & $\pm$~0.7  % 132-Te 228 
 & \textless~2.9~ & $\pm$~0.8  % 132-I 668 
 & \textless~23~ & $\pm$~7  % 133-I 530 
 & \textless~19~ & $\pm$~5  % 134-Cs 605 
 & 18~ & $\pm$~5~$\pm$~5  % 137-Cs 662 
\\ 
\hline

\end{tabular}
\end{table*}

We looked for a correlation between the activities obtained for the fission products and the cosmogenic isotope \nuc{7}{Be} in order to probe the influence of weather effects on the observed fluctuations in activities. For this we calculated the Pearson product-moment correlation coefficient (see, e.g., \citet{Cow98}) between the fission isotope activities and \nuc{7}{Be}. The values obtained for the correlation coefficient between 0.34 and 0.61 indicate a ``medium'' correlation. This suggests that part of the observed fluctuations are due to changing weather conditions that lead to more or less activity passing through the air filters.
 
\begin{figure}[t]
\vspace*{2mm}
\begin{center}
\includegraphics[width=1.0\linewidth, angle=0]{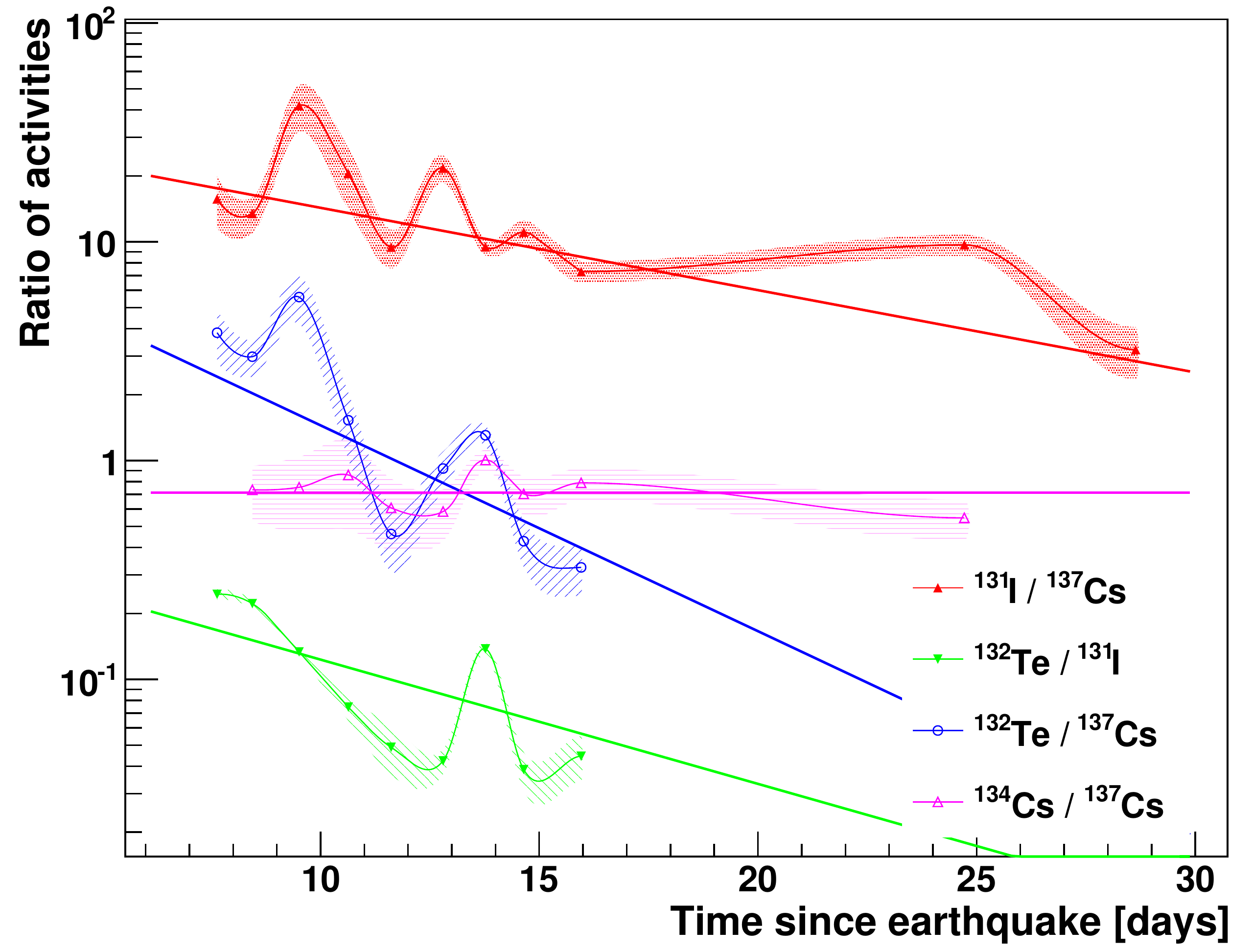}
\end{center}
\caption{\label{activitiesRatios}(Colour online) Ratio of the activities for the isotopes specified in the legend. The error band shows the statistical error only as the systematic errors cancel to a large degree in the ratios. The fits to the ratios as a function of time were performed with fixed decay time constants. For a discussion see text.}
\end{figure}

\begin{figure}[!t]
\vspace*{2mm}
\begin{center}
\begin{tabular}{c}
\includegraphics[width=1.0\linewidth, angle=0]{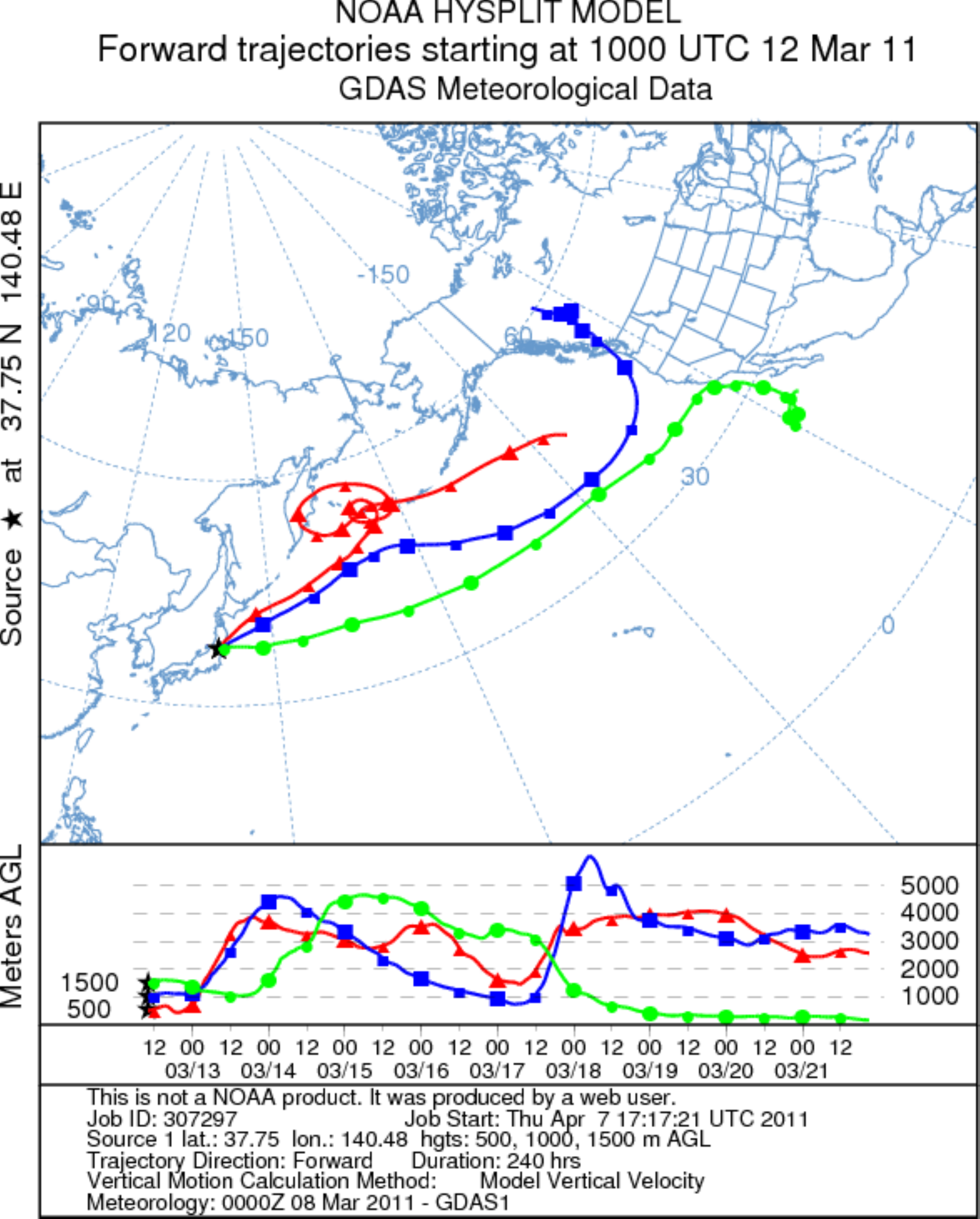} \\
(a) \\
\includegraphics[width=1.0\linewidth, angle=0]{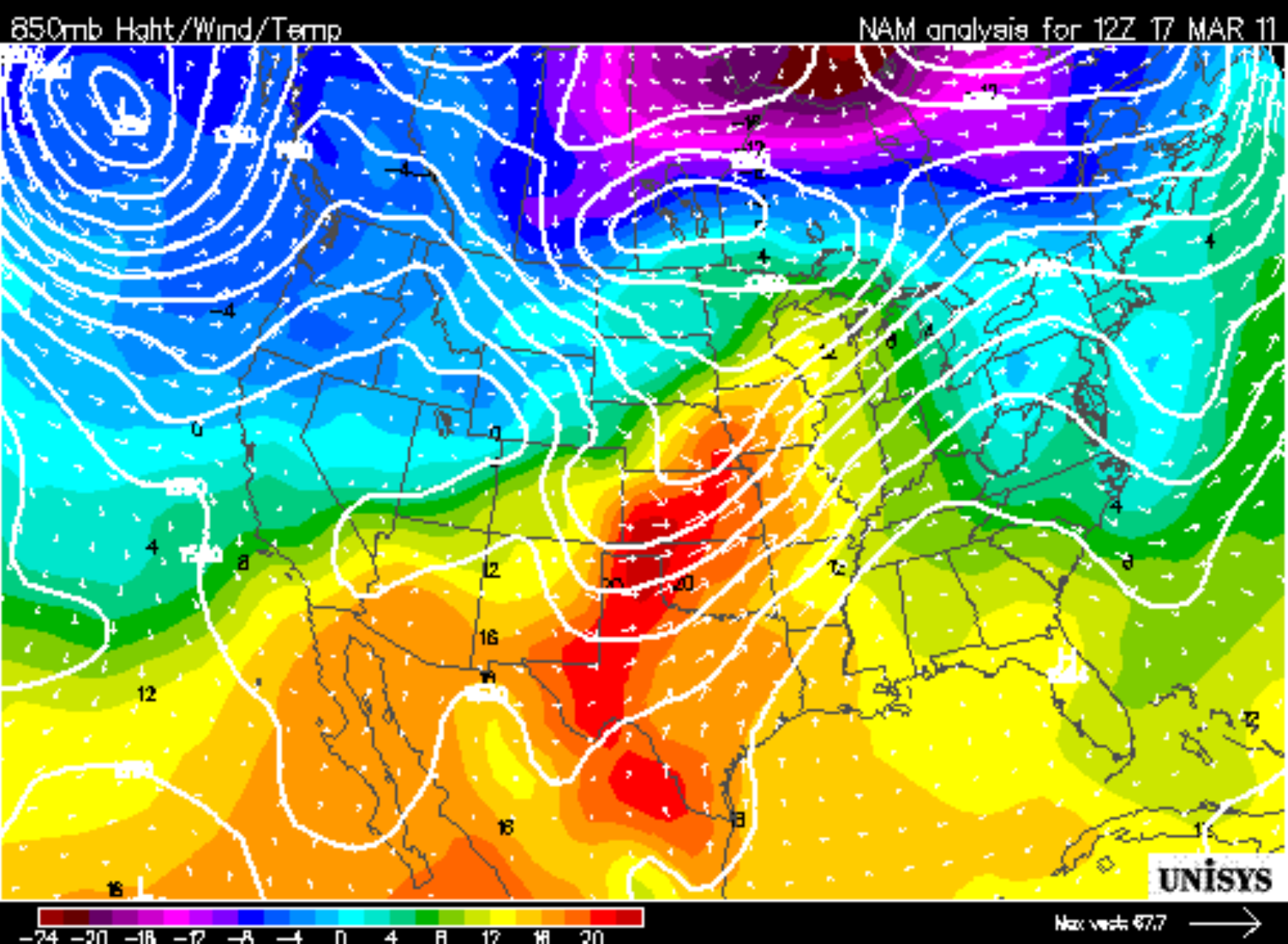} \\
(b) \\
\end{tabular}
\end{center}
\caption{\label{trajectory}(Colour online) Calculated trajectories for airmasses released at the site of the Fukushima Dai-ichi reactor (a) and the 850~mb geopotential heights, wind vectors and temperatures for 17 March 2011, 12 UTC (b). For more details see text.}
\end{figure}

In addition to the activities themselves we also analyzed the ratio of activities for several isotopes. While the activities have a large systematic error due to the different efficiency corrections these errors cancel to a very large degree in the ratios. Figure~\ref{activitiesRatios} shows the ratios for \nuc{131}{I} /  \nuc{137}{Cs}, \nuc{132}{Te} /  \nuc{131}{I}, \nuc{132}{Te} /  \nuc{137}{Cs}, and \nuc{134}{Cs} /  \nuc{137}{Cs}. We show also fits to the ratios $r$ with a simple exponential decay
\begin{equation}
r(t) = r_0 e^{-t \left( \frac{1}{\tau_1}-\frac{1}{\tau_2} \right)}
\end{equation}
where the lifetimes $\tau_i$ have been fixed to the known values of the different isotopes. We excluded the points where only an upper limit for one of the activities was available. While the ratio of the two caesium isotopes is nicely described by this fit (a constant fit in that case due to the large half lives with $\chi^2$/dof = 10.1/8) the other ratios show large deviations ($\chi^2$/dof $\gtrsim $5) and the fit rather follows the general trend of the ratio. This indicates that the release and transport efficiencies for the different elements vary in time leading to ``bursts'' of individual elements during certain days. It has already been alluded above that certainly the transport of \nuc{131}{I} is different from the other radionuclides due to the differences in the measured filter efficiencies. The release and transport of \nuc{131}{I} is also more efficient than that of \nuc{132}{Te} as can be seen by extrapolating the ratio of  \nuc{132}{Te} to  \nuc{131}{I} back to the end of active reactions in the fuel rod (day 0 in Fig.~\ref{activitiesRatios}). Comparing the extracted value of $\sim$0.5 to the ratio of fission yields of $\sim$1.5 \citep{IP} shows that more \nuc{131}{I} was released and/or transported than \nuc{132}{Te}.

In order to assess the transport time of the fission products across the Pacific we performed several model calculations. The trajectories were computed using the NOAA HYSPLIT model \citep{Drax11} using the Global Data Assimilation System (GDAS) meteorological dataset and model-calculated vertical velocities. We used the HYSPLIT model to calculate several hundred trajectories over the time period of interest. The trajectories show a range in transport patterns depending  on the altitude and hour of the start time. Figure~\ref{trajectory} (a) shows three trajectories which exhibit the range of transport pathways. The start time is 12 March 2011, 10 UTC, which is approximately 3 hours after the first reported explosion from Unit 1 \citep{Tepco}, and the trajectories were calculated for three heights in the boundary layer, 500, 1000 and 1500 meters above ground level. The trajectory at 500 meters is caught up in, and lofted by, a cyclonic system over the Bering Sea. The trajectories started at 1000 and 1500 meters are partially lofted by the same system, but do not enter the cyclonic pattern. Instead they are rapidly transported across the Pacific. Upon arrival to the west coast of the U.S. the transport again splits with one arm transported to the north in a cyclonic direction around a second low pressure system located off the coast of Washington state. Figure~\ref{trajectory} (b) shows the 850~mb geopotential heights, wind vectors and temperatures for 17 March 2011, 12 UTC. The trajectory initially started at 1500 meters is transported in the boundary layer towards California.  The northern arm is again lofted by a cold front near Washington state. There were rain showers and cool weather in Western Washington at this time. The strong divergence and precipitation associated with these weather systems most likely significantly reduced the concentrations of radionuclides that were transported. The trajectories support the  notion of transport of the radionuclides from the Japanese boundary layer to the U.S. boundary layer in only 5 to 6 days. This is significantly faster than previously reported trans-Pacific times \citep{Wei07,NRC}, especially considering the radionuclides were released in the boundary layer over Japan and measured in the boundary layer along the U.S. west coast. 
 
Following the nuclear accident, measurements on radiation levels on land and off the coast of Fukushima from several sources were collected and made available by the \citep{MEXT}. Highest reported readings for \nuc{131}{I}  in the region area are in the range of 500-600~Bq\,m$^{-3}$ on March 22nd and 23rd about 25~km south of the plant, although there is an extreme value of 5600~Bq\,m$^{-3}$ reported for March 21st. The Tokyo Electric Power Company (TEPCO) is constantly monitoring \nuc{131}{I} and reported one reading of 4100~Bq\,m$^{-3}$ within about 10 km of the reactor on March 20th. Most other observations by MEXT and TEPCO are less than 10~Bq\,m$^{-3}$. The Japanese Ministry of Defense also reported \nuc{131}{I} values using aircraft samples, with a maximum of 0.46~Bq\,m$^{-3}$ for March 25th in an area 25-30~km to the west. Based on our highest observed value of 4.4~mBq\,m$^{-3}$, we estimate that this air was diluted by a factor of 10$^\mathrm{5}$-10$^\mathrm{6}$ prior to reaching the boundary of the continental U.S. This level of dilution is not surprising given the transport patterns mentioned above.

From the presence and/or absence of certain fission isotopes we can draw several conclusions on their origin:  (i) The value of the ratio of \nuc{134}{Cs} to \nuc{137}{Cs} activities of $\sim$0.7 is indicative of the release of the fission products from a nuclear reactor and not from nuclear weapons \citep{Dev86}. (ii) The presence of the relatively short lived isotopes \nuc{131}{I} and \nuc{132}{Te} shows that the fission products had been released primarily from recently active fuel rods as opposed to spent fuel. (iii) The notable absence of 20.8-h \nuc{133}{I} in our spectra, together with the known steady state activity ratio of \nuc{133}{I} to \nuc{131}{I} of $\sim$2 \citep{Dev86,IP} allows us to put a lower limit on the time between the end of steady state nuclear fuel burning and the arrival of the fission products at our location. A lower limit of 5.7 days (95\% C.L.) was derived from the PH2 measurement to be compared with 6.6 days delay between the earthquake and the end of PH2 filter exposure. The PH4 measurement resulted in a lower limit of 8.1 days (95\% C.L.) for the end of exposure of 8.4 days after the earthquake. Given the modelled transport time of 5-6 days from above this means that the reactor successfully shut down at the time of the earthquake and that nuclear reactions were largely stopped at the time of the release of the fission products. Also we found a small indication for a \nuc{133}{I} peak in our PH2 data set. The peak with 22$\pm$8 counts/hour per day of exposure was not significant enough to pass our 3$\sigma$ cut. The reactor shut down time derived from the insignificant \nuc{133}{I}  peak is 6.8$\pm$0.5 days before the end of PH2 exposure which is statistically compatible with the time of the earthquake. (iv) It is striking that we see only three of the many possible fission product elements. This points to a specific process of release into the atmosphere. The exact process and why it would be selective requires further investigation, but we can speculate that the release of fission products to the atmosphere is the result of evaporation of contaminated steam, in which, e.g., CsI is very soluble. Chernobyl debris, conversely, showed a much broader spectrum of elements \citep{Dev86}, reflecting the direct dispersal of active fuel elements.

\section{Conclusions}
We measured the arrival of airborne fission products from the Fukushima Dai-ichi, Japan nuclear reactor incident in Seattle, WA, USA. The first fission products arrived between 17-18 March about 7 days after the earthquake and in agreement with other reported detections of radionuclides in the western United States \citep{Radnet,Tay,Nor11,Bow11}. Our models of the transport of air masses across the Pacific point to a typical transport time of 5 to 6 days. This agrees with the first reported explosion at Fukushima one day after the earthquake \citep{Tepco}, augmented by additional contributions from subsequent explosions.

The detected fission isotopes are \nuc{131}{I}, \nuc{132}{I}, \nuc{132}{Te}, \nuc{134}{Cs}, and \nuc{137}{Cs} with the highest activity observed for \nuc{131}{I} of 4.4$\pm$1.3~mBq\,m$^{-3}$ on 19-20 March. Our measurements were only sensitive to radionuclides attached to particulate mater and were thus not sensitive to the part of \nuc{131}{I} that is transported in its gaseous form. The presence of the aforementioned isotopes clearly points to the release of the fission products from recently active fuel elements. At the same time, the absence of the short lived isotope \nuc{133}{I} indicates that nuclear reactions must have been stopped successfully at the time of the earthquake as we would otherwise have detected its activity.

This set of measurements over a period of 23 days provide a quantitative basis for further modeling on how radioactive fission products are transported in the atmosphere and will hopefully be included in a global analysis.

\section{Acknowledgments}
We are grateful to J. Orrell and H. S. Miley for the loan of equipment and advice. We also benefited from conversations with M. Savage, J. Gundlach, and B. Taylor. The support by the staff of the physics building of the University of Washington, in particular J. Alferness, proved invaluable. This work has been supported by the US Department of Energy under DE-FG02-97ER41020.

%\bibliographystyle{model2-names}
%\bibliography{bibliography.bib}

\end{document}